\documentstyle[11pt,paspconf,epsf,psfig,twoside]{article}

\newcommand{\Line}[3]{\Ion{#1}{#2}\,$\lambda$\,#3}

\newcommand{\Ion}[2]{#1{\,\scriptsize #2}}

\def\edcomment#1{\iffalse\marginpar{\raggedright\sl#1\/}\else\relax\fi}
\reversemarginpar

\begin{document}
\vspace{-5.5cm}

\title{Observing the Accretion Stream in Polars}

\author{Jens Kube, Boris T. G\"ansicke, Klaus Beuermann}
\affil{Universit\"ats-Sternwarte, Geismarlandstr.~11, D-37083
G\"ottingen, Germany}

\begin{abstract}

We have developed a 3D computer model of polars which includes the accretion
stream, the accretion column, the secondary star, and the white dwarf. Each
component is represented by a fine grid of surface elements. With brightness
values attributed to each surface element, this model generates synthetic
light curves of the various components. As a first application, we present our
simulated light curves of the accretion stream, fitted to high-speed HST
spectroscopy of the eclipsing polar UZ Fornacis by the means of an evolution
strategy.

\end{abstract}

\section{Introduction}

Orbital phase-resolved spectroscopy of the bright emission lines in polars
provides a unique possibility to derive information on the intensity and
velocity distribution along the accretion stream. In eclipsing polars, the
geometric constraints allow a direct transformation of an emission line light
curve into a brightness map of the accretion stream.  We have developed a
computer model of polars which includes 3D grid models of the secondary star,
the white dwarf, the accretion stream, and the column. Each component is
represented by a large number of surface elements, to which brightness values
are attributed. We calculate the projected surface area of each element for
the desired phase interval and sum up the flux contributions of all surface
elements to obtain the total flux emitted in the direction of the
observer. This light curve can then be fitted to observations in order to
derive intensity maps of the components of the polar. Our present work focuses
on the accretion stream, which is the source of the broad emission lines.

\section{Synthetic light curves}

\begin{figure}[tb]
\begin{tabular}{p{5.5cm}p{7cm}}
\raisebox{-5.5cm}{\epsfysize=5.5cm\epsfbox{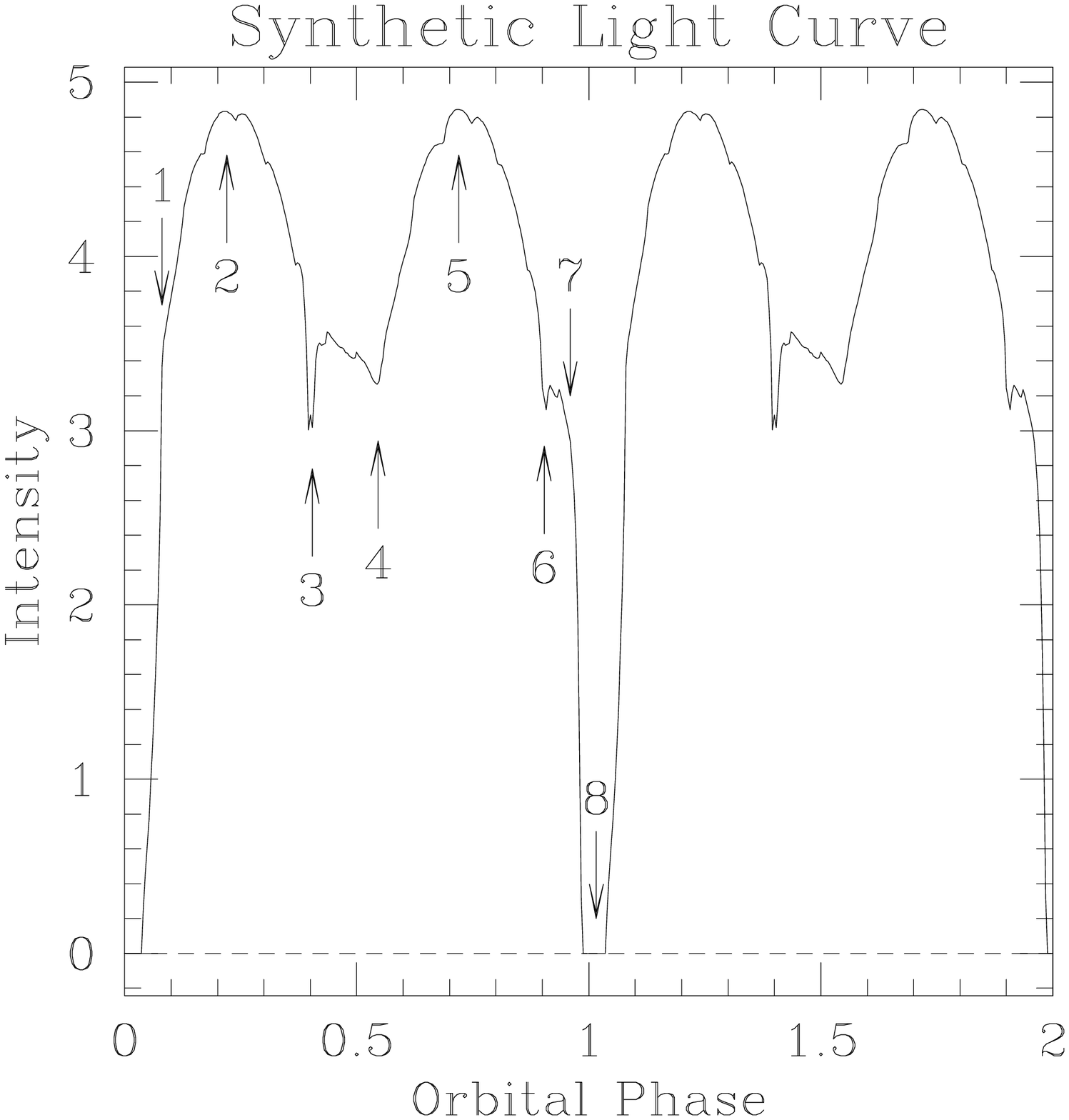}}
&
1\raisebox{-1.2cm}{\epsfxsize=3cm\fbox{\epsfbox{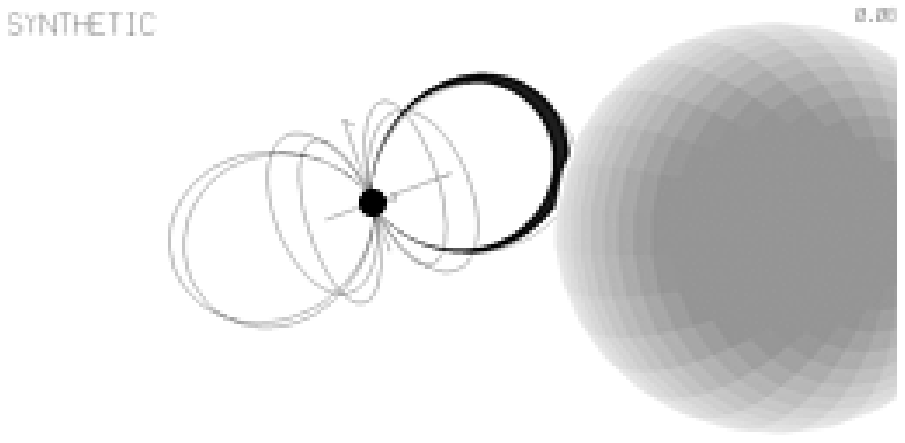}}}
2\raisebox{-1.2cm}{\epsfxsize=3cm\fbox{\epsfbox{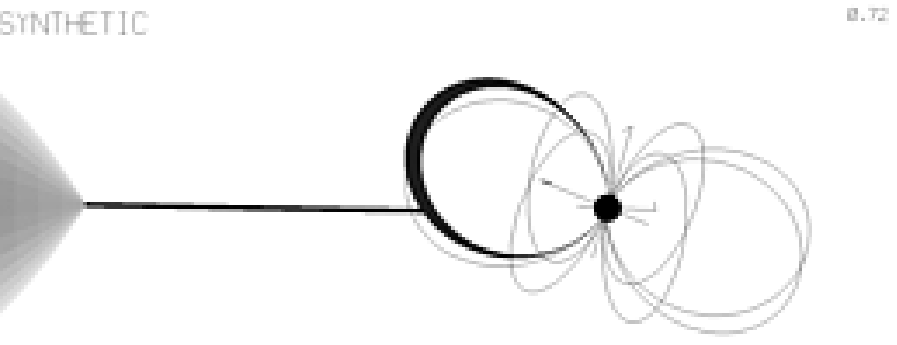}}}

\medskip
3\raisebox{-1.2cm}{\epsfxsize=3cm\fbox{\epsfbox{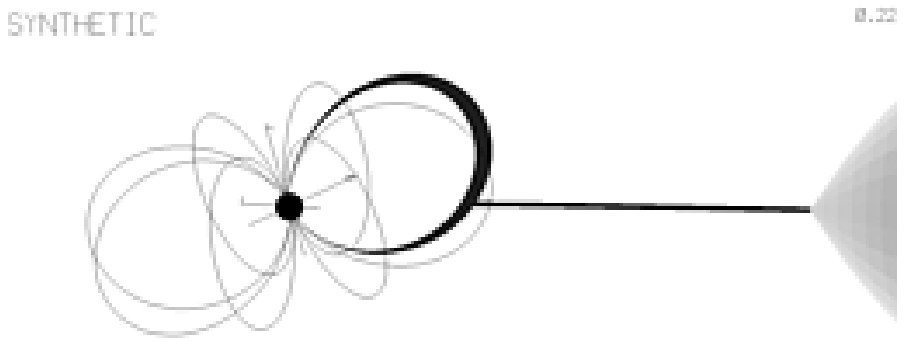}}}
4\raisebox{-1.2cm}{\epsfxsize=3cm\fbox{\epsfbox{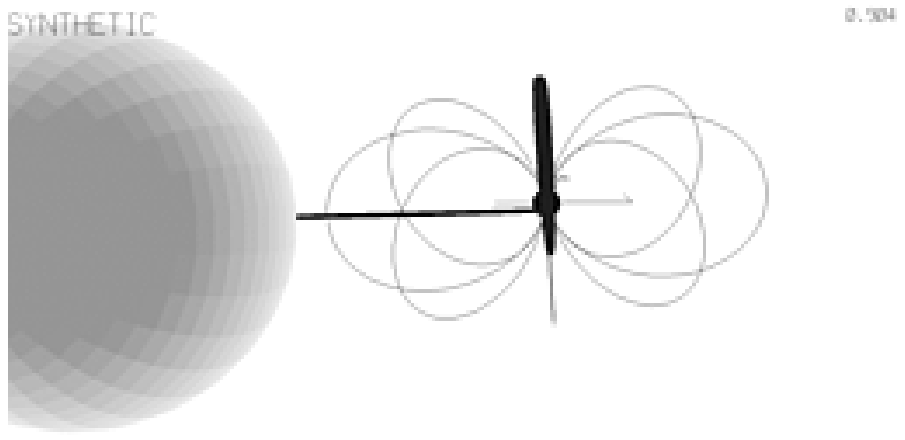}}}

\medskip
5\raisebox{-1.2cm}{\epsfxsize=3cm\fbox{\epsfbox{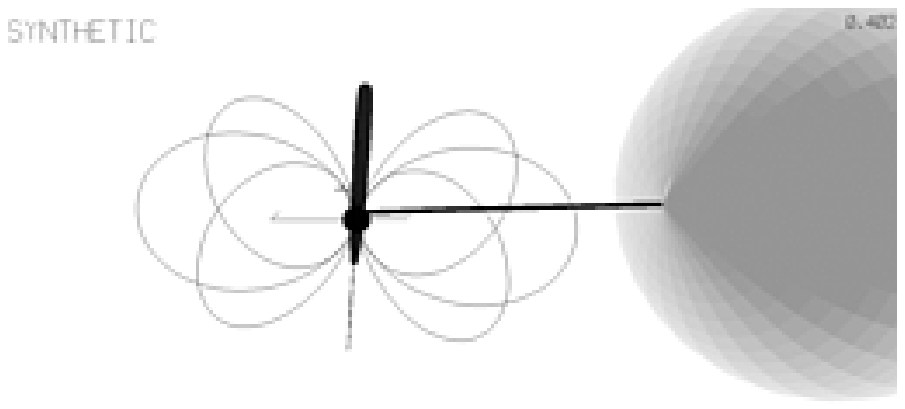}}}
6\raisebox{-1.2cm}{\epsfxsize=3cm\fbox{\epsfbox{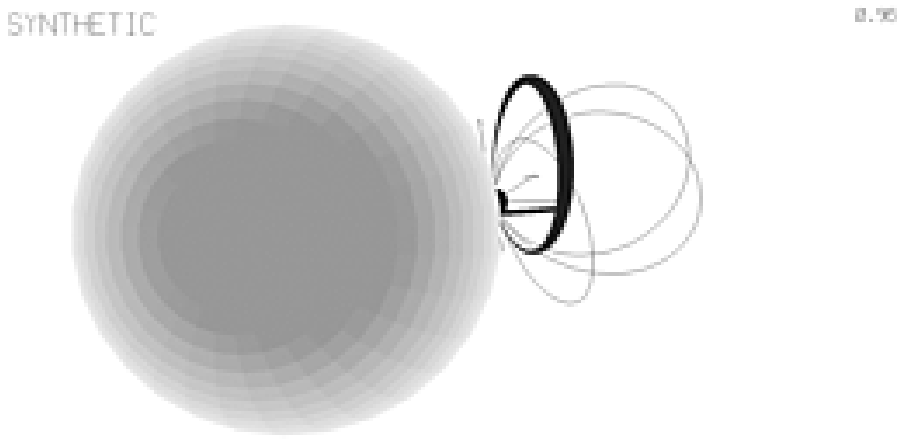}}}

\medskip
7\raisebox{-1.2cm}{\epsfxsize=3cm\fbox{\epsfbox{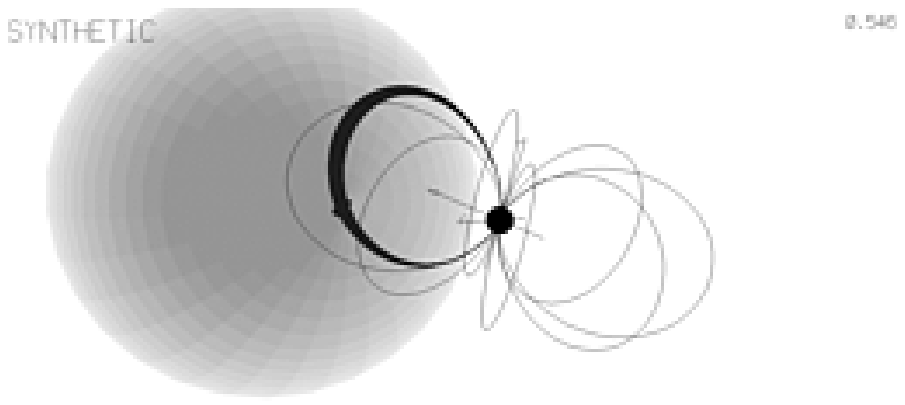}}}
8\raisebox{-1.2cm}{\epsfxsize=3cm\fbox{\epsfbox{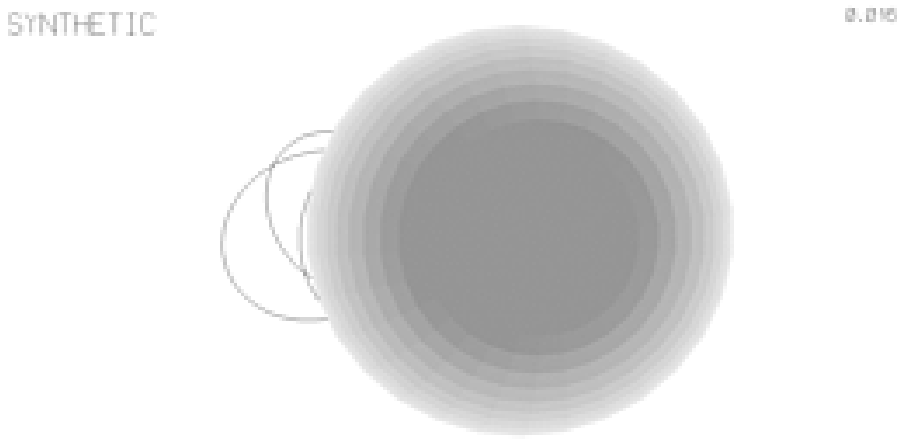}}}
\end{tabular}
\vspace{-0.3cm}
\caption{\small Synthetic light curve of the accretion stream only (left) with
various numbered features and corresponding views of the polar (right)}
\label{fig1}
\end{figure}

In Fig.~\ref{fig1} (left) we show a synthetic light curve for the accretion
stream in a polar at high inclination, $i=88^\circ$. The intensity of the
white dwarf and the secondary star are set to zero; for simplicity, the
accretion stream is assumed to have a uniform surface brightness. Our light
curve shows various features, present also in observed systems (e.g. Schwope
et al., 1997). We follow the features around the orbit (Fig.~\ref{fig1},
right): (1)~The dipole part of the accretion stream has completely reappeared
after the eclipse. (2)~The projected area of the accretion stream is
maximal. (3)~The white dwarf causes a partial eclipse of the accretion
stream. (4)~A secondary minimum occurs, when the projected area of the
ballistic part of the accretion stream is minimal. (5)~As in (2), the
projected area of the accretion stream is maximal. (6)~A dip comes from the
self-eclipse of the dipole part of the stream. (7)~The main eclipse begins. At
first, the stream sections near the $L_1$-point disappear, followed by the
white dwarf and the dipole stream; at last, the stream sections near the
stagnation region are eclipsed. (8)~The stream is eclipsed completely. Note
that the center of the eclipse eclipse of the stream is offset from the center
of the eclipse of the white dwarf.

Schwope et al.~(1997) presented an observed light curve of the broad-line
emission in HU~Aqr which represents the emission of the accretion stream and
shows the same double-hump shape as our calculated light curve in Fig.~1.

\nopagebreak
\section{Geometry}\label{geometry}

\begin{figure}[tb]
\begin{center}
\vspace{-0.6cm}
\begin{tabular}{p{6.5cm}p{5cm}}
\raisebox{-6cm}{\epsfxsize=6.5cm\epsfbox{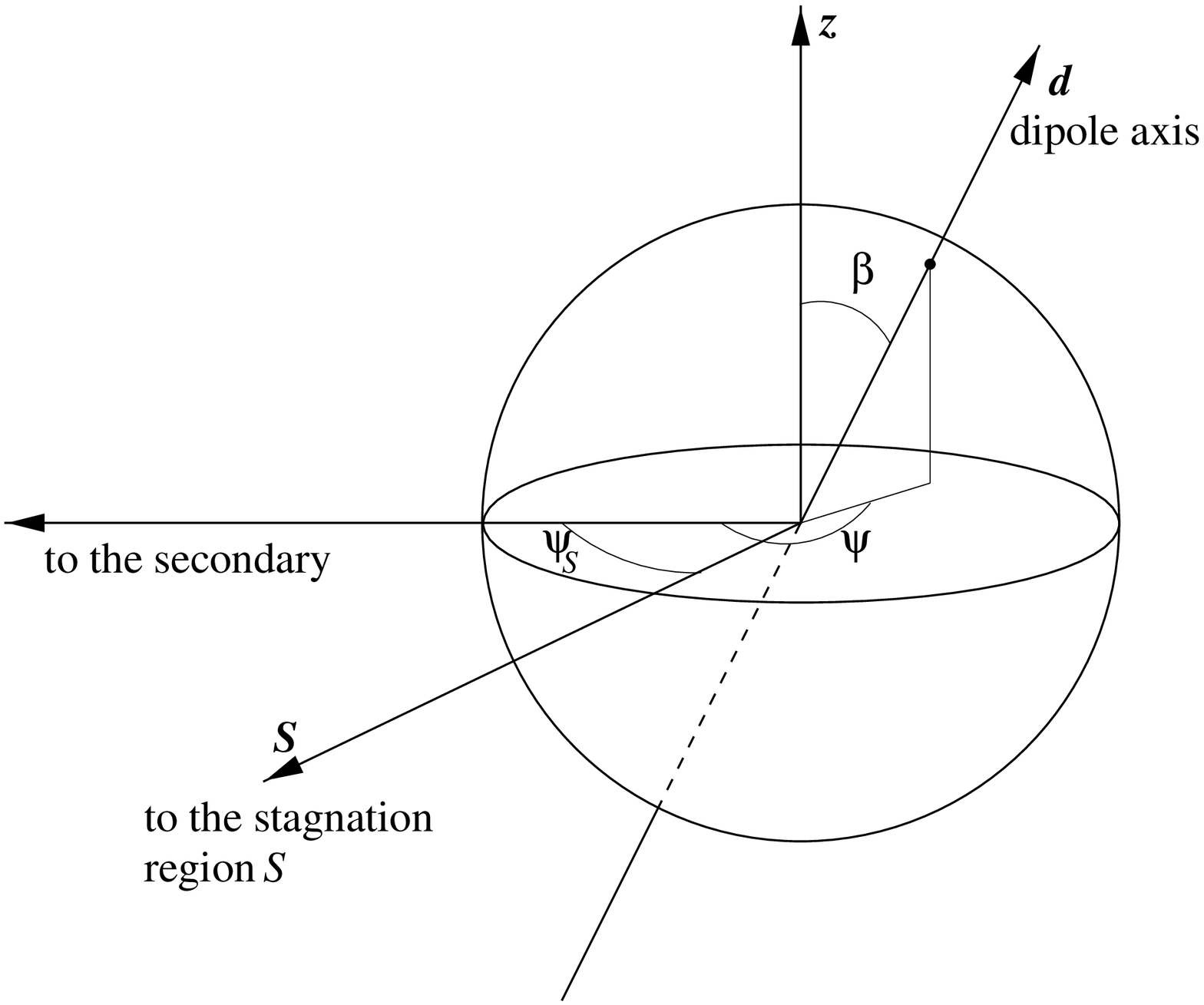}}
&
\medskip
a)\raisebox{-2.8cm}{\epsfysize=3cm\fbox{\epsfbox{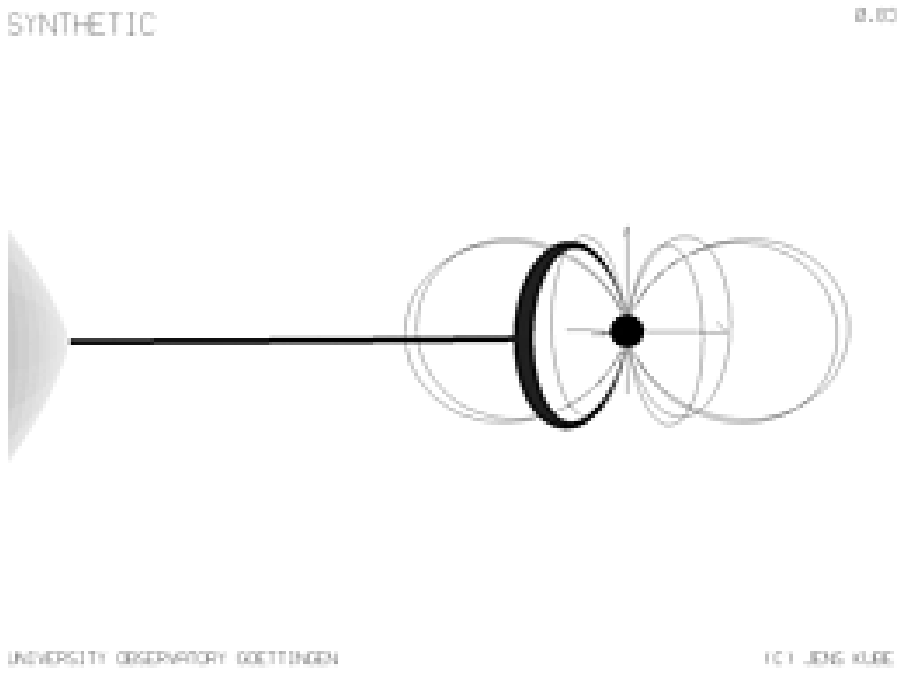}}}

\medskip
b)\raisebox{-2.8cm}{\epsfysize=3cm\fbox{\epsfbox{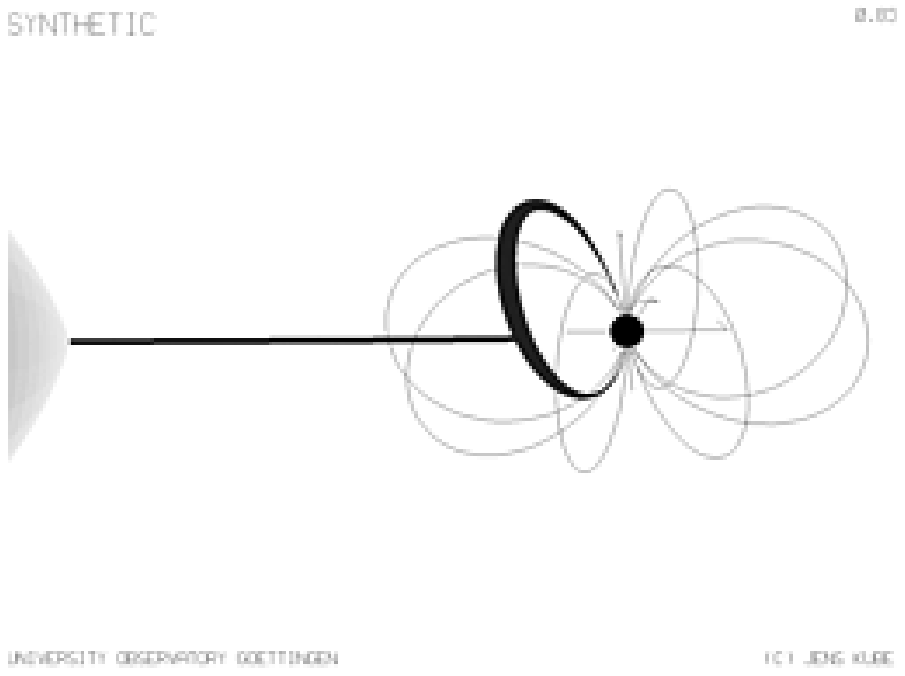}}}

\end{tabular}
\end{center}
\vspace{-0.6cm}
\caption{\small Geometry of the magnetic dipole: definition (left) and examples
(right) (a)+(b) $\Psi_S=35^\circ$, (a) tilt $\beta=20^\circ$, azimuth
$\Psi=75^\circ$, (b) $\beta=0^\circ$, $\Psi=45^\circ$}
\label{fig2}
\end{figure}

The accretion stream geometry depends strongly on the tilt and azimuth of the
dipole axis, characterized by the angles $\beta$ and $\Psi$ in
Fig.~\ref{fig2}. A third characteristic angle is the azimuth of the stagnation
region, $\Psi_S$. These three parameters describe the geometry of the accretion
stream under the following assumptions: (1)~The width of the stagnation region
is given by the dispersion of the free-fall trajectories emerging with a
thermal distribution from the $L_1$-point (Flannery, 1975). (2)~Magnetically
funneled accretion occurs along the same field line for the stream above and
below the orbital plane.  (3)~The center of the dipole lies in the center of
the white dwarf, i.e. there is no dipole offset.

\section{Application: UZ Fornacis}

UZ\,For was observed with HST in June 1992;  a detailed description of
the data is given by Stockman \& Schmidt (1996). We summarize here
only the relevant points. 
\begin{figure}[tb]

\begin{center}
\begin{tabular}{p{5cm}p{5cm}}
\epsfysize=5cm\epsfbox{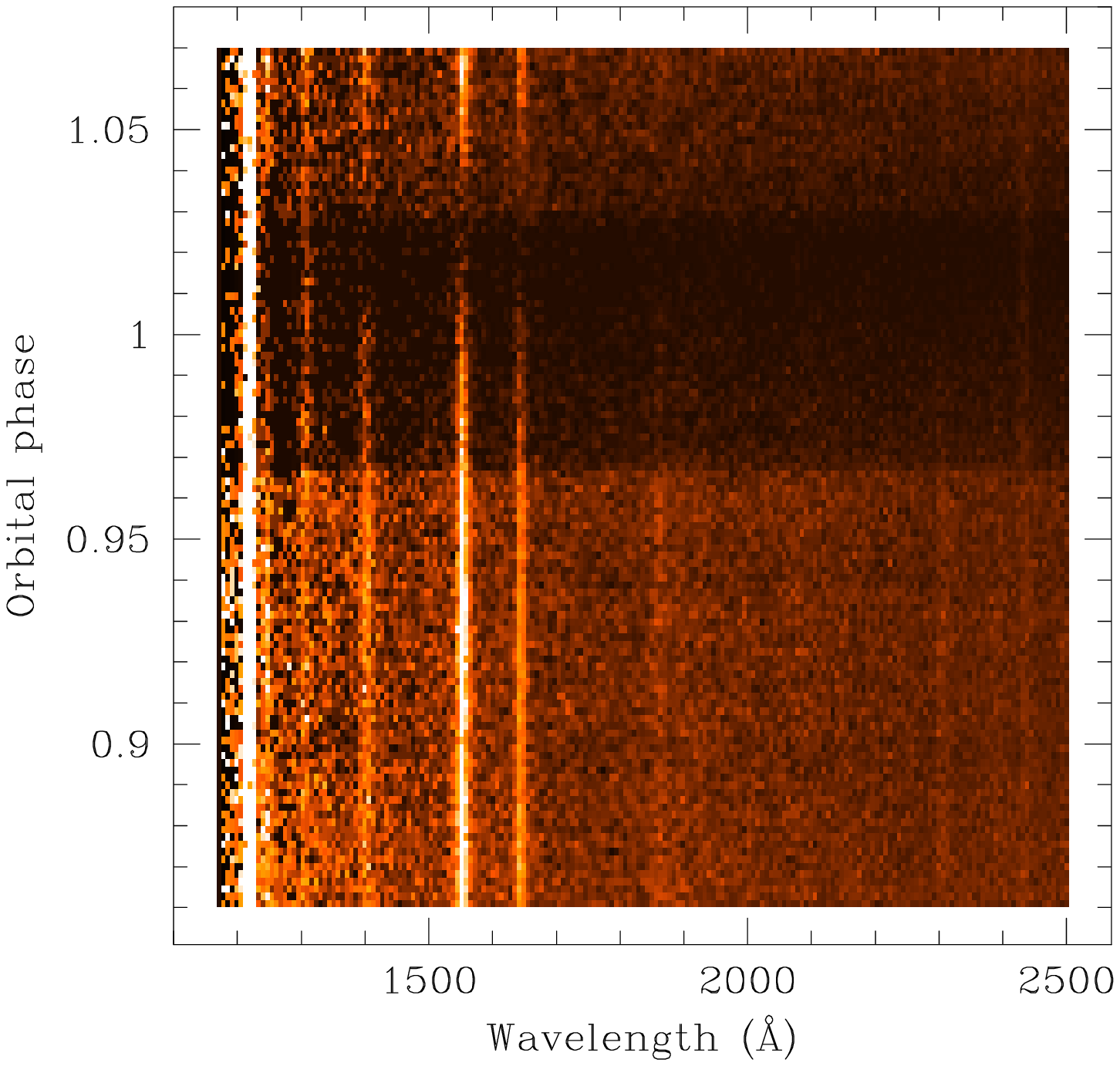}
&
\epsfysize=5cm\epsfbox{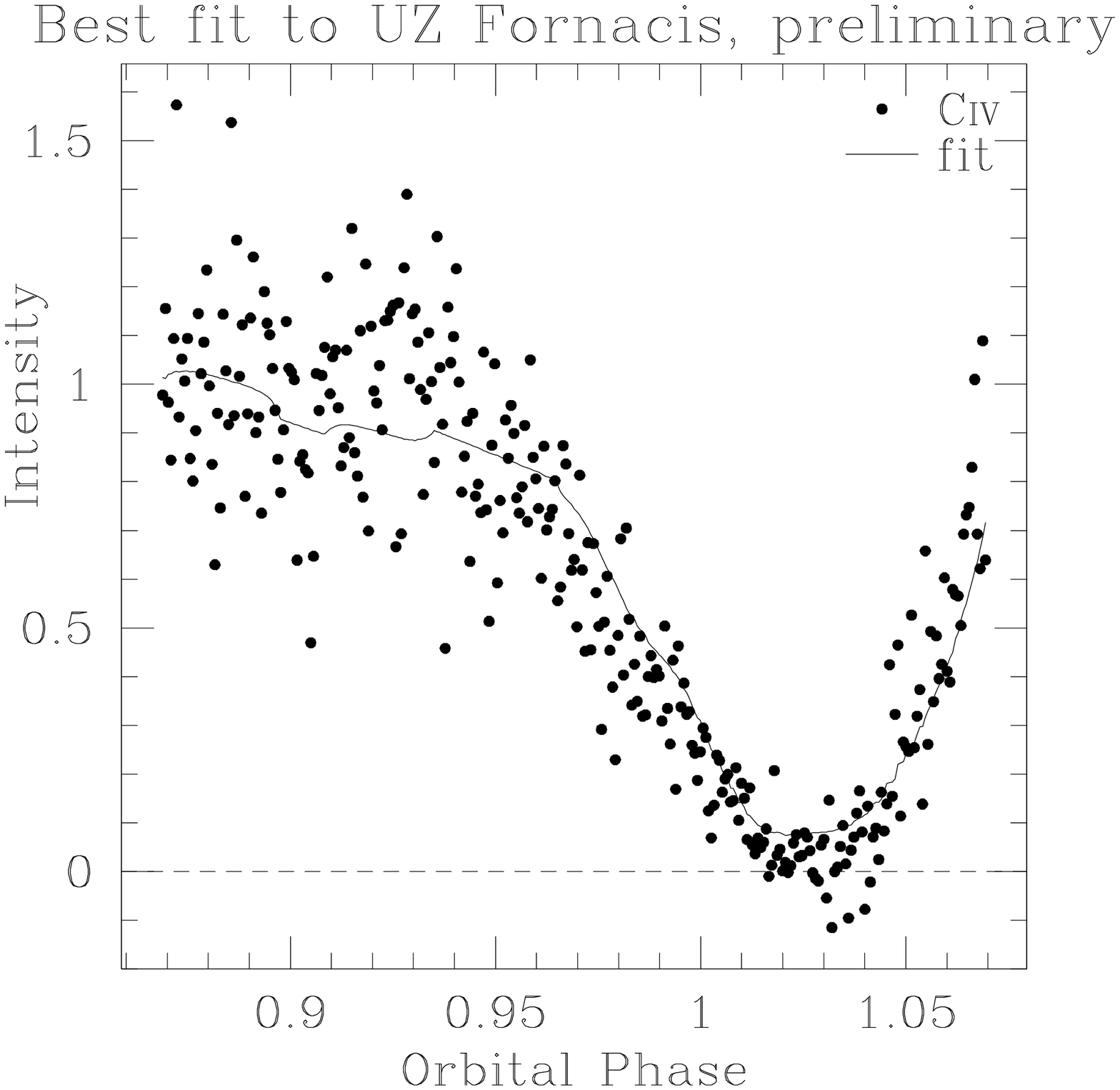}
\end{tabular}
\end{center}
\vspace{-0.6cm}
\caption{\small Raw data (left) and extracted light curve with fit (right)}
\label{fig3}
\end{figure}
Fast FOS/G160L spectroscopy with a time resolution of 1.6914\,s was obtained,
covering two entire eclipses in the phase interval $\phi=0.87\dots1.07$. The
spectra cover the range 1180--2500\,\AA\ with a $\mbox{FWHM
resolution of }\approx7$\,\AA.  We have retrieved these data from the HST archive
at STScI, calibrated with the standard reduction pipeline. The mid-exposure
times of the individual spectra were converted into binary orbital phases
using the ephemeris of Warren et al. (1995). The two individual observations
were then co-added (Fig.~\ref{fig3}, left). From this trailed spectrum, we
extracted the light curve of \Line{C}{IV}{1550} (Fig.~\ref{fig3}, right).

We use the system parameters from Bailey \& Cropper (1991): $q=M_1/M_2=5$,
$M=M_1+M_2=0.85\,M_\odot$, $i=81^\circ$, $P=126.5\mathrm{\,min}$. The optical
light curve in Bailey (1995) shows two eclipse steps which can be explained by
two hot spots on the white dwarf which subsequently dissapear behind the limb
of the secondary star. From that light curve we can measure the timing of the
eclipse events with an accuracy of $\Delta\phi=5\cdot10^{-4}$. The ingress of
the spot on the lower hemisphere occurs at $\phi=0.9685$, its egress at
$\phi=1.0310$; for the spot on the upper hemisphere, ingress is at
$\phi=0.9725$ and egress at $\phi=1.0260$

In addition to the ``basic parameters'' mentioned above we derive the following
\emph{preliminary} dipole configuration from the hot spot eclipse timings:
$\beta=15^\circ$,
$\Psi=-25^\circ$,
$\Psi_S=33^\circ$

Using these parameters, we fit the \Ion{C}{IV} light curve. Fit parameters are
the $\approx 200$ brightness values along the stream. The brightness
distribution is optimized by an evolution strategy, which is the favourable
choice for large numbers of fit parameters. The fit converges after
approximately 300 iterations having calculated roughly 30\,000 light curves in
total. In order to reduce the number of degrees of freedom, we have 
implemented a smoothing algorithm based on maximum entropy.

Our best fit is shown in Fig.~\ref{fig3} (right). This preliminary fit assumes
that accretion to both poles occurs along the same field line. This assumption
turns out to be consistent with the position of the lower main accretion spot,
but not simultaneously with that of the upper secondary spot. Dropping the
assumption of accretion onto both poles on the same field line introduces an
additional degree of freedom.  The detailed 3D brightness distribution will
neccessarily change with any additional assumption made and is not shown
here. Independent of the remaining uncertainties, however, our fit implies
that the accretion stream is brightest near the stagnation region where it
couples onto the magnetic field.

\section{Summary}

Our new model allows light curve synthesis of CVs and tomographical
reconstruction of brightness maps by the means of evolution fit strategies. As
a first result, we have found that the stagnation region in the magnetosphere
of the polar UZ~For is a prominent source of emission lines.

\end{document}